# Application of the DRS4 Chip for GHz Waveform Digitizing Circuit[*]

YANG Hai-Bo(杨海波) [1,2]   SU Hong(苏弘) [1;1)]   KONG Jie(孔洁) [1]   CHENG Ke(成科) [1,3]
CHEN Jin-Da(陈金达) [1]   DU Cheng-Ming(杜成明) [1,2]   ZHANG Jing-Zhe(张惊蛰) [1,2]

1 Institute of Modern Physics, Chinese Academy of Sciences, Lanzhou 730000, China

2 University of Chinese Academy of Sciences, Beijing 100049, China

3. Northwest Normal University, Lanzhou 730070, China

**Abstact:**   At present, fast waveform digitizing circuit is more and more employed in modern physics experiments for processing the signals from an array detector. A new fast waveform sampling digitizing circuit developed by us is presented in this paper. Different with the traditional waveform digitizing circuit constructed with analog to digital converter(ADC) or time to digital converter(TDC), it is developed based on domino ring sampler(DRS), a switched capacitor array(SCA) chip. A DRS4 chip is used as a core device in our circuit, which has a fast sampling rate up to five gigabit samples per second (GSPS). The circuit has advantages of high resolution, low cost, low power dissipation, high channel density and small size. The quite satisfactory results are acquired by the preliminary performance test of this circuit board. Eight channels can be provided by one board, which has a 1-volt input dynamic range for each channel. The circuit linearity is better than 0.1%, the noise is less than 0.5 mV (root mean square, RMS), and its time resolution is about 50ps. The several boards can be cascaded to construct a multi-board system. The good performances make the circuit board to be used not only for physics experiments, but also for other applications.

## 1   Introduction

The maximum information of the pulse from detector can be acquired by experimenter by digitizing a fast signal directly, for example, the drift time and pulse area can be easily obtained from the waveform [1]. Traditionally, flash ADC is used for digitization [2-3], however such systems suffer from sampling rates in the range about from 50 to 250 MHz with 10 or 12 bits resolution, low channel densities, huge power consumption and are usually expensive. Due to development of component technology now, an alternative approach is to utilize switched-capacitor arrays (SCA) [4-6]. The input signal are sampled and stored in a series of capacitors at high sample rates under the control of a shift register, and digitized with a commercial ADC operating at lower sample rate.

In our design, a DRS4 [7], the fourth version chip of DRS, is chosen. Because

[*] Supported by National Natural Science Foundation of China(11305233), Specific Fund Research Based on Large-scale Science Instrument Facilities of China(2011YQ12009604)

1)   E-mail:suhong@impcas.ac.cn

the domino wave runs continuously in a ring, the chip is called domino ring sampler (DRS). DRS from Paul Scherrer Institute (PSI) in Switzerland is a typical SCA [8]. Because of the high channel density of the DRS system, it becomes affordable to be used for waveform digitization in experiments which ADC/TCDs are used currently. This technique has been developed for particle physics applications. At present, it has been realized using switched capacitor circuits, for example MAGIC [9] and MEG [10]. It also can be useful for other applications, such as PET scanners and portable oscilloscopes.

A prototype of waveform sampling readout board based on DRS4 chip is developed, and preliminary test is implemented. DRS4 waveform digitizing board records the input signal with a high sample rate between 0.7 and 5 GSPS. Its gain can be adjusted to match different kinds of detector signals. The input dynamic range of the waveform digitizing board is about 66 dB at a sample rate of 5 GS/s, and the time resolution is about 50ps.

The key parameters of the waveform digitizer are as follows:
1) Input bandwidth ≥ 650MHz;
2) 1 V input dynamic range;
3) 16 bit DAC calibration;
4) Up to 5 GSPS sampling rate;
5) Digital trigger logic and external trigger implemented in the FPGA;
6) 14 bit ADC resolution;
7) Serial DRS4 readout;
8) Readout multiplexed and speed up to 33 MHz;
9) Board to board communication.

## 2  Design of the waveform digitizing board

2.1  The DRS4 technology and operation

DRS4 chip is fabricated in 0.25 mm 1P5M MMC process and is a radiation hard SCA. DRS4 has up to 5GHz sampling speed, 9 differential input channels per chip, 11.5 bit vertical resolution, 4ps timing resolution [11]. The waveform is stored in 1024 sampling cells per channel, and can be readout after sampling with a commercial ADC at 33 MHz sampling rate. The range of sampling rate is from the GHz to the MHz. The time stretch ratio (TSR) is given as shown in Eq.(1). Dead time is described by Eq.(2).

$$TSR \equiv \delta t_s / \delta t_d \quad (1)$$

$$\text{Dead time} = \text{Sampling Window} * TSR \quad (2)$$

Where, $\delta t_s$ is the sampling interval of DRS4, $\delta t_d$ is the sampling interval of readout ADC.

For example, typical values: $\delta t_s$ =0.5ns (2 GSPS), $\delta t_d$ =30ns (33MHz), sampling window=100ns, the result of TSR=60, dead time=100ns*60=6μs.

The main parameters of DRS4 chip are shown in Table 1.

Table 1. DRS4 chip main parameters.

| type | parameter value |
|---|---|
| power supply | single 2.5 V |
| sampling speed | 700 MSPS to 5 GSPS |
| analog channel | 8+1 |
| storage depth | 1024 cells |
| high SNR: | ~ 69 dB |
| low noise | ~ 0.35 mV |
| analog outputs | multiplexed or parallel |
| channel or chip cascading | yes |
| readout mode | full or region of interest |
| inputs bandwidth | 950 MHz |

The DRS4 chip consists of an on-chip a series of inverters generating a sampling frequency in the GHz range. A sampling signal propagates through these inverters freely (domino principle). There is an analog voltage supplied to the chip, then, transmission gates between these inverters make the sampling frequency controllable in a wide range. The frequency of the domino wave is stabilized by a phase locked loop (PLL) and defined by an external reference clock. The differential input signal is stored in 150 fF capacitors. The domino wave runs continuously in a circular fashion, but domino wave can be stopped at any sampling cell by a trigger signal. The trigger signal stops the domino wave, and then analog contents of the sampling cells are read out under the control of a shift register, and digitized by an external ADC simultaneously. Fig. 1 shows the simplified schematic of the chip.

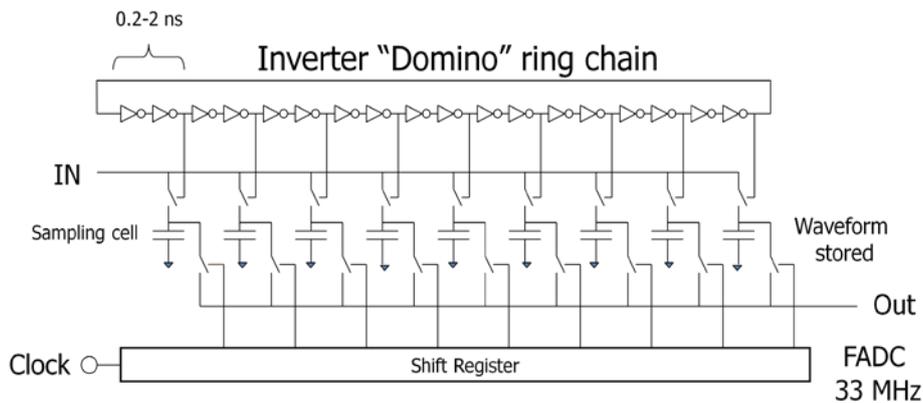

Fig.1. Simplified schematic of the DRS4 chip.

The DRS4 chip has three kinds of waveform readout modes [10]. A feature of the DRS4 chip is its transparent mode, whereby this mode, the analog input to the DRS4 chip is digitized and applied to the analog output at the same time. In the full readout mode, all 1024 sampling cells are read out consecutively starting from cell 0 with 1024 clock cycles for external digitization. In the region of interest readout mode, it can be read only a subset of all sampling cells, and the dead time is reduced. The DRS4 chip has a sampling depth of 1024 cells per channel. For the applications to obtain deeper sampling depth, it supports cascading two or more channels on one chip, and even several DRS4 chips can be daisy-chained.

## 2.2 Circuit design

### 2.2.1 Overview of the waveform sampling digitizing board

The overview of the waveform sampling digitizing board is shown in Fig. 2. The circuit consists of a DRS4 chip, an ADC chip, a FPGA and other devices. The digitizing board has eight analog channels. The DRS4 chip is capable of sampling differential input channels, so the single-end inputs terminated by 50-Ohm need to be converted into differential signals, which is achieved by active buffers chosen. The on-board 16-bit DAC generates reference voltages to measure the offsets of all sampling cells for calibration, and generates offset voltages for DRS4 and buffers. The calibration information of board is stored in an electrically erasable programmable read-only memory (EEPROM). The input signal has a 1V maximum amplitude, and AC coupled mode is adopted at inputs. The data in DRS4 is read by a field programmable gate array (FPGA) and a 14-bit commercial ADC. Controlling digitizing board, receiving outputs of ADC and communicating with a personal computer (PC), are implemented entirely by a FPGA. A low-end FPGA from Xilinx Spartan 3 families is selected as the FPGA in central control unit (CCU) [12]. The data and commands are transmitted between a PC and digitizing board via universal serial bus (USB), the connection between the digitizing board and PC is implemented with a USB controller. The data transfer rates of the USB bus is over 20 MB/sec. A comparator is used in each analog input channel for generating a triggering signal. The one input of comparator comes from the non-inverting input of differential inputs of DRS4, and another one, a programmable level is set as a reference. After comparing, a trigger signal is generated by the comparator. These trigger signals from all comparators can be combined into "AND" or "OR" logic in FPGA, and then a trigger is generated by FPGA. Once the trigger is effective, the sampling in DRS chip is stopped and the information stored in SCA are digitized with an external ADC at a sampling rate of 33MHz. A prototype of waveform sampling digitizing board, as shown in Fig.3, is realized, which can be used to construct a small data acquisition system.

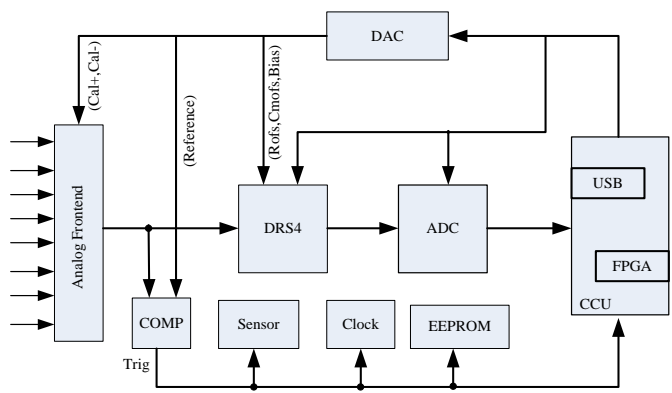

Fig. 2. Block diagram of a waveform sampling digitizing board.

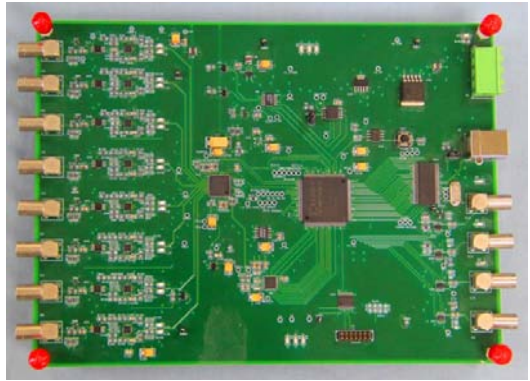

Fig.3 The prototype of waveform sampling digitizing board.

2.2.2 Analog Frontend

Fig. 4 illustrates the schematic of the analog frontend. As the inputs of DRS4 are differential signals, a wideband, fully-differential operational amplifier THS4508, from Texas Instruments®, is chosen as differential driver [13], the bandwidth of the amplifier reaches 2 GHz. The current given by the THS4508 for driving a DRS4 is about 1mA. An analog switch ADG901, from Analog Devices®, is placed at the front end of THS4508, which is dc-coupled with the input of THS4508. When the DRS4 is calibrated, this switch is switched off, the input of the circuit is isolated. AC coupling mode is applied to the inputs of DRS4 and ADG901, the advantage is that it can protect the input devices and simplify the processing for common mode levels. As the NMOS transistors at the inputs of DRS4 show a nonlinear behavior, the linearity becomes a litter worse at the portions near the rails, it is recommended to be operated within the entire linear range. The input baseline is generated by a DAC to map input signals, which can apply an individual DC offset to the differential input lines. Because the analog frontend is used for processing high speed analog signals, some detailed methods are adopted for designing a printed circuit board (PCB), such as matched impedance, separate power supply and proper termination. The analog frontend circuit is simulated by PSpice, the three dB bandwidth of the analog frontend is about one GHz when the load of circuit is a 10pF capacitance.

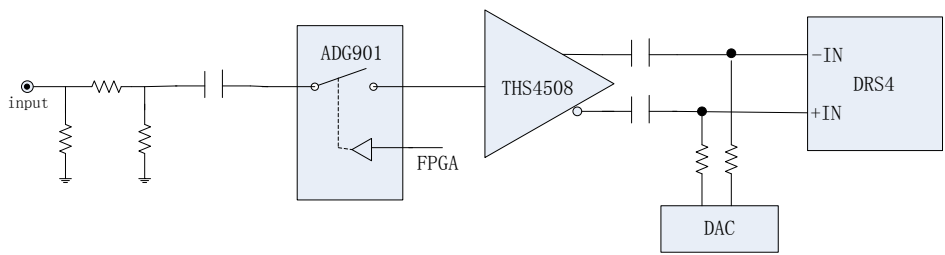

Fig. 4. Schematic of the analog frontend.

2.2.3 Digitization

All DRS4 channels are readout one by one for digitization, an ADC is required. There is a very strict phase relationship between the DRS4 readout clock and the

ADC clock. The phase relationship of two clocks must be fixed and about 10ps jitter tolerated for the best linearity, and a delay circuit is designed for generating a phase shifted clock with a low jitter. The signal to noise ratio (SNR) of DRS4 is 66 dB after offset correction. In order to get the optimal performances of the waveform sampling readout board, a 14-bit ADC from Analog Devices is used.

### 2.2.4  Control voltages and calibration

In order to let DRS4 chip operating normally, certain configuring voltages are needed, such as DC offset, BIAS, ROFS and O-OFS. BIAS and O-OFS can be set internally by the DRS4 itself, also, can be set externally by supplying a bias voltage. In our design, an external 16-bit DAC is used to generate a bias voltage connected to these control lines. The bias voltage can be fine-tuned to compensate for variations of the chip. This DAC also provides threshold voltage to inverting input(input-) of comparator, while one event comes, this comparator can output a trigger signal to FPGA. These control voltages are shown in Fig.5. A 0Vcalibration voltage offered by16-bit DAC is sent to all DRS4 inputs for the voltage calibration. Because more driving current is required for driving inputs of DRS4, a low noise Op Amp AD8605[14] is employed as a buffer after the DAC. Calibration test point is measured and an offset voltage and gain are evaluated. As parameters of transistors in the chip gradually changes normally, the timing calibration is necessary. In order to reduce nonlinearity, we need measure the delay of each cell. An internal 240 MHz clock is sampled by one channel of DRS4, and the deviation between the expected period and the measured period is used to determine the effective width of each cell. This design ensures the jitter of timing calibration is less than 20 ps. The values of calibration voltage and calibration timing data are saved in an EEPROM.

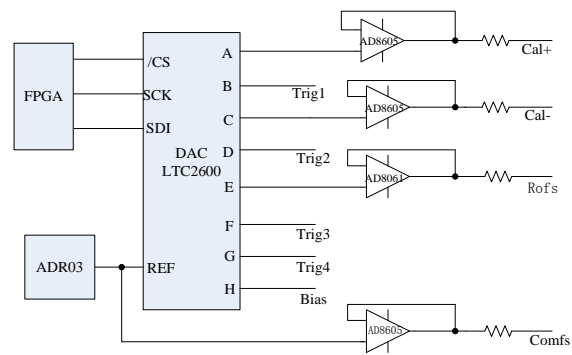

Fig. 5.   The circuit for producing control voltages.

### 2.2.5  Multi-board configuration

The waveform sampling digitizing board can be used for a detector that has multiple detection cells. Several waveform sampling digitizing boards can be cascaded, and a multi-board system, supplying multiple analog input channels, is constructed. The multi-board configuration is shown in Fig.6. There are four control inputs on each board, Trigger IN, Trigger OUT, Clock IN and Clock OUT. The port "Trigger IN" accepts an external trigger when a hit event is coming, which much like a trigger of oscilloscope. The port "Trigger OUT" sends the pulse with a fixed width to the "Trigger IN" port on next board. The "Clock IN/OUT" ports allow a better synchronization among different boards for multi-board configuration. Both the trigger and the clock signals are passed by a daisy-chain mode from the master board to the slave board. In this system, more than eight input channels are read out for, each event. The data from different boards can be distinguished by the board number.

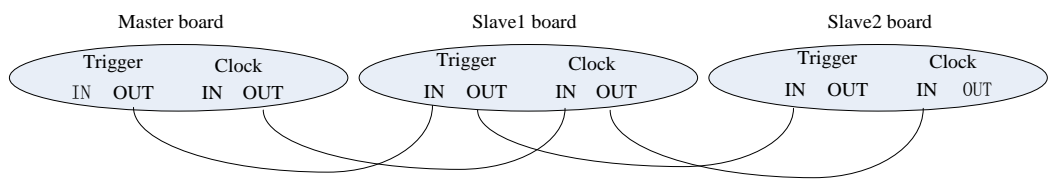

Fig.6.    Scheme of multi-board configuration.

## 3    Test and Results

The preliminary test of the prototype board is implemented, and the results of measurement are discussed in this section.

### 3.1    Baseline noise

The root mean square (RMS) noise on baseline is measured before offset correction, and also measured after offset correction. Because of some factors caused by integrated circuit (IC) manufacture technology, each sampling cell in SCA has a fixed residual voltage (i.e. DC offset). The DC offset basically is fixed for the sampling cell, but DC offset has some differences among the different sampling cells. If DC offset is not corrected, it can cause nonlinear distortion to the reconstruction of the waveform. Therefore, DC offset of each sampling cell is needed to be calibrated and removed from the sampling result. In the calibration, a 0V voltage, generated by the16-bit DAC, is given to each input channel of DRS4. Statistical analysis of the sampled results is performed for all the sampling cells. Fig. 7 shows a 0 V DC signal is sampled at 5 GSPS sampling rate before offset and gain correction, which shows a noise level of 7.3 mV RMS. Fig. 8 shows a 0 V DC signal is sampled at 5 GSPS sampling rate after offset and gain correction, the noise level is reduced significantly to about 0.5mV RMS. So "fixed pattern" offset error of 7.3 mV RMS can be reduced to 0.5 mV RMS by offset correction implemented in FPGA. Thus, for $1V_{peak-peak}$ dynamic input range, the SNR is about 66 dB (1 V linear range / 0. 5mV).

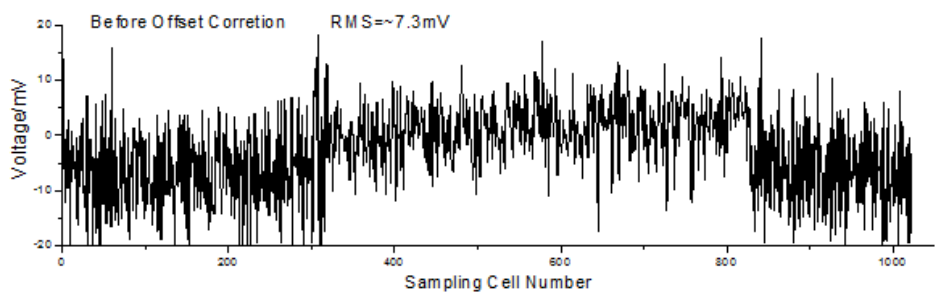

Fig. 7.    0 V DC signal sampled at 5 GSPS before offset correction.

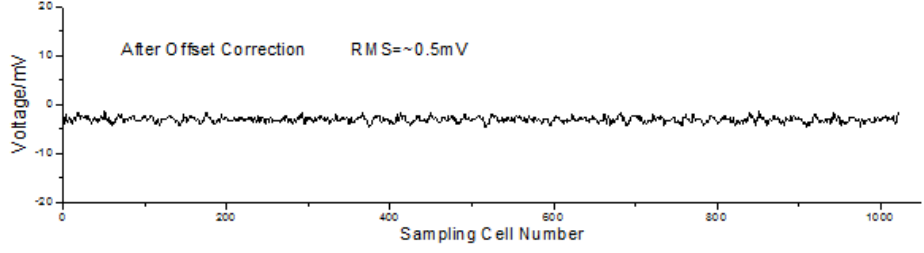

Fig. 8. 0 V DC signal sampled at 5 GSPS after offset correction.

Another measure method of the SNR is to obtain the effective number of bits (ENOB) [15]. For an input sine wave signal with the specified frequency and amplitude, ENOB is defined by the Eq. (3).

$$ENOB = \log_2\left(\frac{FSR}{NAD\sqrt{12}}\right) \approx N - \log_2\left(\frac{NAD}{\varepsilon_Q}\right) \quad (3)$$

where, N is the number of bits digitized, FSR is the full-scale range of the recorder, NAD is the noise and distortion, $\varepsilon_Q$ is the RMS ideal quantization error.

In the test, a sine wave signal is generated, which has a peak-peak value of about 0.45 V, the frequency is 20MHz. Fig.9 gives the typical result, NAD=1.31mV, $\varepsilon_Q$=2V ×2^-14/√12=3.524e-5. These values is brought into Eq. (3), then:

ENOB=14-$\log_2$ (1.31/0.035)=14-5.2=8.8 bit.

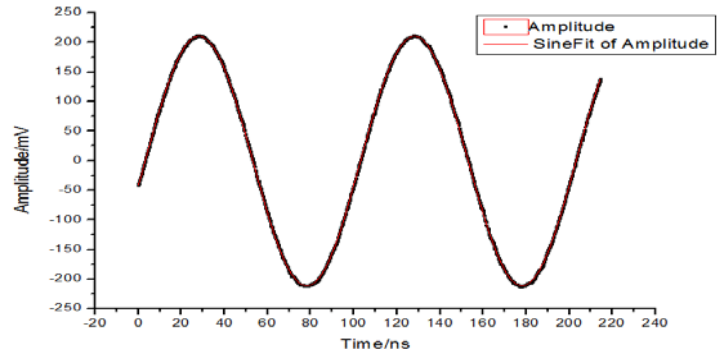

Fig.9. Reconstruction of 10 MHz sine wave signal sampled at 5 GSPS.

3.2 Amplitude nonlinearity

The amplitude nonlinearity is measured by using the pulse signals with 1 MHz frequency generated by AFG3252 [16]. The amplitude of input signal is adjusted from - 400 mV to +400 mV, the step adjusted each time is 100 mV. Input signals are sampled with 5GHz sampling rate, and the amplitudes of the signals are read out with 33MHz sampling rate. The linearity of one channel of the digitizing board is shown in Fig. 10, and the amplitude nonlinearity is about 0.1%. Fig.11 shows output residuals deviating from the linear fit values after offset and gain calibration, the error points are the maximum deviation values.

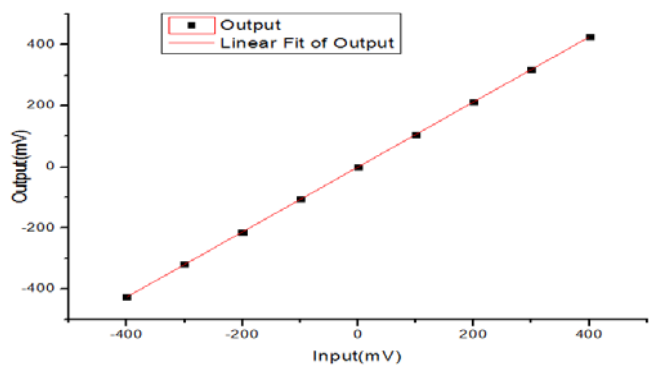

Fig.10 . Analog output vs. analog input.

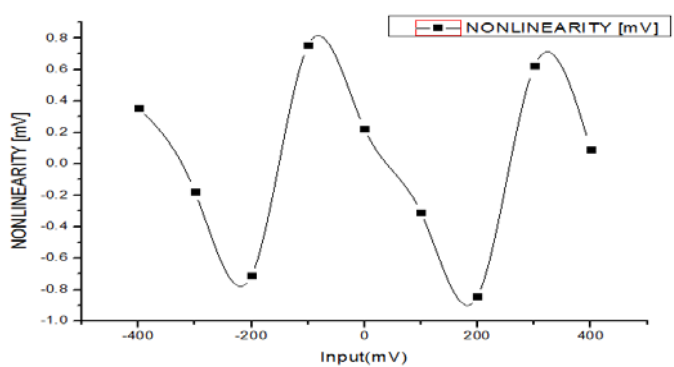

Fig.11. Residuals deviating from the fit value, typical nonlinearity after offset and gain calibration.

A typical detector signals are shown in Fig. 12, which is sampled by the DRS4 chip with the 2GHz sampling rate. An experimental test is implemented using two identical detector modules. Each detector module consists of an 5.2cm×5.2cm×1.5cm LYSO scintillation crystal matrices coupled to the Hamamatsu H8500 PMT wrapped in Teflon. Two LYSO scintillation detector units are separated 16cm away from each other, and 22Na source is placed between them as positron source [17]. Each H8500 has sixty-four anode signals, the 64 anode signals are divided by an external voltage divider board that consists of resistor chains. The sixty-four anode signals can be sent out individually via two x and two y electrodes. The outputs of voltage divider board offer energy information, and the dynode 12 of H8500 PMT offers time information. Results on energy and timing information are shown as an example in Fig. 10, including two time signals and four energy signals.

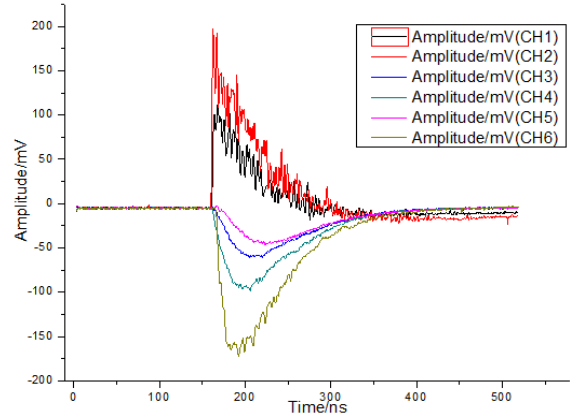

Fig.12. Waveforms of the two time signals and four energy signals produced by DRS4.

3.3 Precision of timing measurement

In the DRS4 chip, because of the possible variation of delay interval of inverter gate, the "delay interval width" of each sampling cell is different, and therefore needs to be measured and calibrated. An internal 240 MHz clock is sampled by one channel with 5.12GHz sampling rate, the delay interval of each sampling cell is obtained and which is shown in Fig13.

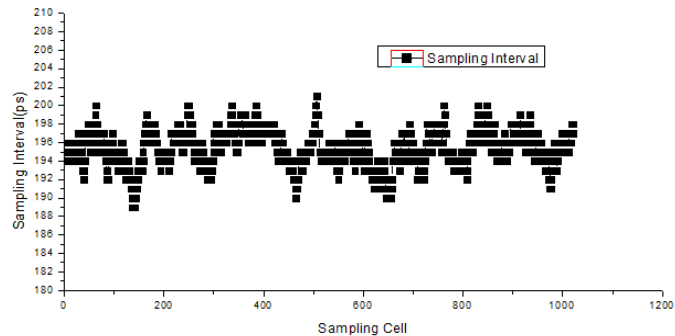

Fig.13. DRS4 delay interval of each sampling cell.

The precision of timing measurement is studied by a delay time measurement based on a cable delay method. A pulse signal from the signal generator AFG3252 [16] is split and fed into two channels with a fixed cable delay between them. A fixed 9.28ns delay time is offered by one cable. Tests are performed, and tow pulses output from two channels of DRS4 is shown in Fig.14. The distribution of delay interval tested is shown in Fig.15, and the sigma, width of delay time spectrum, is 52ps, which includes contribution of time jitter of the signal generator itself.

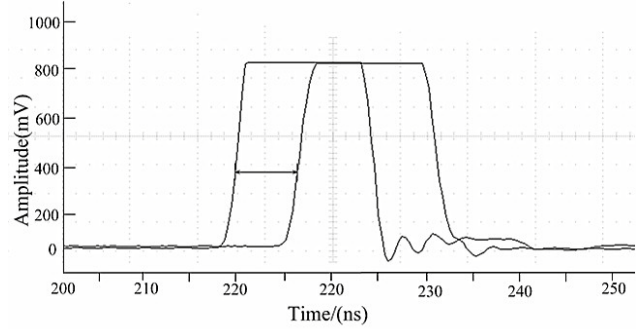

Fig.14. The pulse of two-channel signals.

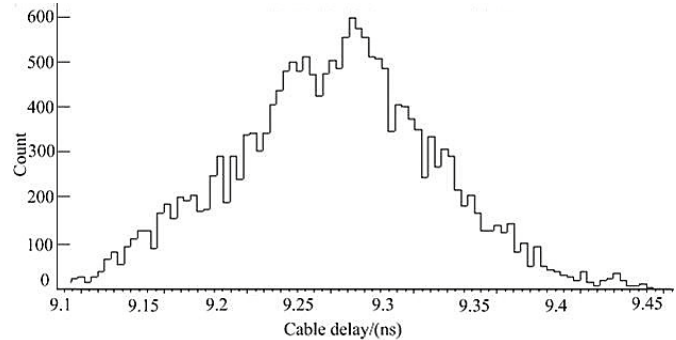

Fig.15. Distribution of a cable delay time tested.

## 4  Conclusions

A new flexible waveform sampling readout digitizing board, based on DRS4 chip, is realized and tested. The sampling rate for an input signal can be changed from 700MHz to 5 GSPS, and the sampling rate of ADC for readout digitization is 33 MHz. The bandwidth (-3 dB) of the digitizing board is about 700MHz, the input dynamic range is one volt. The SNR is about 66 dB, and the time resolution is approximately 50ps. A high speed sampling based on SCA is combined with an ADC with lower sampling rate but high precision, which makes the system with the features of high resolution, low cost, low power dissipation, high channel density and small size. It can replace some digitizing system based on flash ADC for waveform digitization. It is very useful for a waveform digitizing system with multiple input channels. The waveform sampling readout digitizing board can be employed to construct data acquisition system for other applications as well.

**Acknowledgements**

This work is supported by the National Natural Science Foundation of China (No. 11305233) and Specific Fund of National key scientific instrument and equipment development project (No. 2011YQ12009604).

**References**

1. Wang J H, Zhao L, Feng C Q, et al. Nucl. Sci. Technol, 2012, 23(2): 109-113
2. Dhawan S, Hughes V W, Kawall D, et al. Nucl. Instrum. Methods A, 2000, 450(2): 391–398
3. Kornilov N V, Khriatchkov V A, Dunaev M, et al. Nucl. Instrum. Methods A, 2003, 497: 467–478
4. Haller G M, Wooley B A. IEEE Trans Nucl Sci, 1994,41:1203–1207
5. Kleinfelder S. IEEE Trans. Nucl. Sci., 1990, 37(3): 1230–1236
6. Delagnes E, Degerli Y, Goret P, et al. Nucl. Instrum. Methods A, 2006, 567: 21–26
7. Ritt, S. IEEE Nuclear Science Conference Record ,2008, 1512-1515
8. Ritt S, Dinapoli R, Hartmann U. Nucl. Instr. and Methods A, 2010, 623(1): 486-488
9. Baixeras C. arXiv preprint astro-ph/0403180, 2004
10. Mori T. PSI R-99-05 MEG Experiment Proposal, Paul Scherrer Institute, 1999
11. Paul Scherrer Institute, DRS4 Datasheet, 2009
12. Xilinx Corporation. Application Note, 774 (v.1.2), 2006
13. Texas Instruments Incorporated, THS4508 Datasheet,2008
14. Analog Devices, AD8605 Datasheet, 2009
15. IEEE Standards , IEEE Std 1057-2007, Revision of IEEE 1057–1994, 2007
16. Tektronix Corporation. AFG3252 User Guide, 2008
17. Chen J D, Xu H S, Hu Z G, et al. Chinese Physics C, 2011, 35(1): 61